\documentclass[twocolumn,groupedaddress,prb]{revtex4}
\usepackage{amssymb}
\usepackage{amsmath}
\usepackage{graphicx}
\usepackage{color}

\begin{document}

\title{Canonical sampling through velocity-rescaling}
\author{Giovanni Bussi}
\email{gbussi@ethz.ch}
\author{Davide Donadio}
\author{Michele Parrinello}
\affiliation{Computational Science, Department of Chemistry and Applied Biosciences,
ETH Z\"urich, USI Campus, Via Giuseppe Buffi 13, CH-6900 Lugano, Switzerland}

\begin{abstract}
We present a new molecular dynamics algorithm for sampling the canonical
distribution. In this approach the velocities of all the  particles
are rescaled by a properly chosen  random factor. The algorithm
is formally justified and it is shown that, in spite of its stochastic
nature, a quantity can still be defined that remains constant during the
evolution. In numerical applications this quantity can be used to measure
the accuracy of the sampling.
We illustrate the properties of this new method on Lennard-Jones and
TIP4P water models in the solid and liquid phases.
Its performance is excellent and largely independent on
the thermostat parameter also with regard to the dynamic properties.
\end{abstract}

\maketitle

\section{Introduction}

Controlling the temperature and assessing the quality of the trajectories
generated are crucial issues in any molecular dynamics simulation.\cite{alle-tild87book,frenk-smit02book}
Let us first recall that in conventional molecular dynamics the microcanonical ensemble NVE
is generated due to the conservation laws of Hamilton's equations.
In this ensemble the number of particles $N$, the volume $V$ and the energy $E$ are kept constant.
In the early days of molecular dynamics the temperature was controlled by
rescaling the velocities until the system was equilibrated at the target
temperature. Energy conservation was also closely monitored in order to
check that the correct NVE ensemble was being sampled and as a
way of choosing the integration time-step. Furthermore, energy conservation
provided a convenient tool for controlling that the code was free from obvious
bugs.

Only in 1980, in a landmark paper,\cite{ande80jcp} Andersen suggested that
ensembles other than the microcanonical one could be generated in a molecular dynamics run
in order to better mimic the experimental conditions.
Here we only discuss his proposal for generating the canonical ensemble
NVT, in which the temperature $T$ rather than the
energy $E$ is fixed. Andersen's prescription was rather simple: during the simulation
a particle is chosen randomly and its velocity extracted from
the appropriate Maxwell distribution. While formally correct, Andersen's
thermostat did not become popular. A supposedly poor efficiency was to
blame, as well as the fact that discontinuities in the trajectories were
introduced. However its major drawback was probably the fact that one had to deal
with an algorithm without the comforting notion of a conserved quantity on which to rely.
Another type of stochastic
dynamics which leads to a canonical distribution is Langevin dynamics.\cite{schn-stol78prb} Such a
dynamics is not often used because it does not have an associated conserved
quantity, the integration time-step is difficult to control and
the trajectories lose their physical meaning
unless the friction coefficient is small.
For similar reasons, an algorithm\cite{heye83cp} which is close to the simplified version
of ours discussed in Section~\ref{sec-rescale}, has been even less popular.
Inspired by the extended Lagrangian approach
introduced in Andersen's paper to control the external pressure,
Nos\'{e} introduced his by now famous thermostat.\cite{nose84jcp}
Differently from Andersen's thermostat, the latter allowed
to control the temperature without using random numbers.
Furthermore, associated with Nos\'{e}'s dynamics there was
a conserved quantity. Thus it is not surprising that Nos\'{e}'s thermostat is widely used,
especially in the equivalent form suggested by Hoover.\cite{hoov85pra}
However, Nos\'{e} thermostat can exhibit non ergodic behavior. In order to
compensate for this shortcoming the introduction of chains of thermostats was
suggested,\cite{mart92jcp} but this spoils the beauty and simplicity of the theory and
needs extra tuning.

Alternative thermostats were suggested such as that of Evans\cite{evan-morr84cpc}
in which the
total kinetic energy is kept strictly constant. This leads to a well defined
ensemble, the so-called isokinetic ensemble, which however cannot be
experimentally realized. Another very popular thermostat is that of
Berendsen.\cite{bere+84jcp} In
this approach Hamilton's equations are supplemented by a first order equation
for the kinetic energy, whose driving force is the difference between the
instantaneous kinetic energy and its target value. Berendsen's thermostat is stable,
simple to implement and physically appealing, however it has no conserved
quantity and is not associated to a well defined ensemble, except in
limiting cases. In spite of this, it is rather widely used.

In this paper we propose a new method for controlling the temperature that
removes many of the difficulties mentioned above. Our method is an extension of the
Berendsen thermostat to which a properly constructed random force is added,
so as to enforce the correct distribution for the kinetic energy. A 
relaxation time of the thermostat can be chosen such that the dynamic
trajectories are not significantly affected. We show that it leads to the
correct canonical distribution and that there exists a unified scheme in
which Berendsen's, Nos{\'e}'s and our thermostat can be formulated.
A remarkable result is that a quantity can be defined which is constant
and plays a role similar to that of the energy in the microcanonical ensemble.
Namely, it can be used to verify how much our numerical procedure generates
configurations that belong to the desired NVT ensemble
and to provide a guideline for the choice of the integration time-step.
It must be
mentioned that all the algorithms presented here are extremely easy to
implement.

In Section \ref{sec-theory} we shall first present a simpler version of our algorithm which is an
extension of the time honored velocity-rescaling. Later we shall describe
its more general formulation, followed by a theoretical analysis of the new
approach, a comparison with other schemes and a discussion of the
errors deriving from the integration with a finite time-step.
The following Sections \ref{sec-applications} and \ref{sect-conclusion}
are devoted to numerical checks of the theory 
and to a final discussion, respectively.

\section{Theory}
\label{sec-theory}

\subsection{A canonical velocity-rescaling thermostat}
\label{sec-rescale}

In its simplest formulation, the velocity-rescaling method consists in
multiplying the velocities of all the particles by the same factor $\alpha
$, calculated by enforcing the total kinetic energy $K$ to be equal to
the average kinetic energy at the target temperature $\bar{K}=\frac{N_{f}}{%
2\beta }$, where $N_{f}$ is the number of degrees of freedom and $\beta $ is
the inverse temperature. 
Thus, the rescaling factor $\alpha$ for the velocities is obtained as 
\begin{equation}
\alpha=\sqrt{\frac{\bar{K}}{K}}.
\end{equation}%
Since the same factor is used for all the particles, there is neither an effect on
constrained bond lengths nor on the center of mass motion. This operation is
usually performed at a predetermined frequency during equilibration, or when the
kinetic energy exceeds the limits of an interval centered around the target
value. The sampled ensemble is not explicitely known but, since
in the thermodynamic limit the average properties do not depend on the ensemble
chosen, even this very simple algorithm can be used to produce useful
results. However, for small systems or when the observables of
interest are dependent on the fluctuations rather than on the averages,
this method cannot be used. Moreover, it is questionable to assume
that this algorithm can be safely combined with other methods which require
canonical sampling, such as replica-exchange molecular dynamics.\cite%
{sugi-okam99cpl}

We propose to modify the way the rescaling factor is calculated, so as
to enforce a canonical distribution for the kinetic energy. Instead of
forcing the kinetic energy to be exactly equal to $\bar{K}$, we select its
target value $K_{t}$ with a stochastic procedure aimed at obtaining the
desired ensemble.
To this effect we evaluate the velocity-rescaling factor as
\begin{equation}
\alpha=\sqrt{\frac{K_{t}}{K}},
\end{equation}%
where $K_{t}$ is drawn from the canonical equilibrium distribution for the
kinetic energy: 
\begin{equation}
\bar{P}(K_{t})~dK_{t}\propto K_{t}^{\left(\frac{N_f}{2}-1\right)}e^{-\beta K_{t}}~dK_{t}.
\label{kinetic-energy-distribution}
\end{equation}%
This is equivalent to the method proposed by Heyes,\cite{heye83cp}
where one enforces the distribution in 
Eq.~(\ref{kinetic-energy-distribution}) by a Monte Carlo procedure.
Between rescalings we evolve the system using Hamilton's equations. The
number of integration time-steps can be fixed or randomly varied. Both the
Hamiltonian evolution and the velocity-rescaling leave a canonical
probability distribution unaltered. Under the condition that the
Hamiltonian evolution is ergodic in the microcanonical ensemble, it follows
that our method samples the canonical ensemble.\cite{mano-deem99jcp} More
precisely, Hamilton's equations sample a phase-space surface with fixed center
of mass and, for non-periodic systems, zero angular momentum. Since the
rescaling procedure does not change these quantities, our algorithm samples
only the corresponding slice of the canonical ensemble. We shall neglect
this latter effect here and in the following as we have implicitly done in
Eq.~(\ref{kinetic-energy-distribution}).

\subsection{A more elaborate approach}

The procedure described above is very simple but disturbs considerably the
velocities of the particles. In fact, each time the rescaling is applied the
moduli of the velocities will exhibit a fast fluctuation with relative
magnitude $\sqrt{1/N_f}$.
Thus, we propose a smoother approach in which the rescaling
procedure is distributed among a number of time-steps. This new
scheme is somehow related to what previously described in the same way as
the Berendsen thermostat is related to standard velocity
rescaling.

First we note that it is not necessary to draw $K_{t}$ from the distribution in
Eq.~(\ref{kinetic-energy-distribution}) at each time-step:
the only requirement is that the random
changes in the kinetic energy leave a canonical distribution unchanged. In
particular, the choice of $K_{t}$ can be based on the previous value of $K$
so as to obtain a smoother evolution. We propose a general way of doing this
by applying the following prescriptions:

\begin{enumerate}
\item Evolve the system for a single time-step with Hamilton's
equations, using a time-reversible area-preserving integrator such as
velocity Verlet.\cite{velocityV}
\item Calculate the kinetic energy.
\item Evolve the kinetic energy for a time corresponding to a single
time-step using an auxiliary continuous stochastic dynamics.
\item Rescale the velocities so as to enforce this new value of the kinetic
energy.
\end{enumerate}

The choice of the stochastic dynamics has some degree of arbitrariness,
the only constraint being that it has to leave the canonical distribution in
Eq.~(\ref{kinetic-energy-distribution}) invariant. Here we choose this
dynamics by imposing that it is described by a first-order differential
equation in $K$. Since the auxiliary dynamics on $K$ is one-dimensional, its
associated Fokker-Planck equation\cite{gard03book} must exhibit a
zero-current solution. It can be shown that under these conditions the most
general form is 
\begin{equation}
dK=\left( D(K)\frac{\partial \log \bar{P}}{\partial K}+\frac{\partial D(K)}{%
\partial K}\right)~dt+\sqrt{2D(K)}~dW,
\end{equation}%
where $D(K)$ is an arbitrary positive definite function of $K$,
$dW$ a Wiener noise, and we are using the Itoh convention.\cite{gard03book}
Inserting the distribution of Eq.~(\ref{kinetic-energy-distribution})
in this equation one finds
\begin{multline}
dK=\left( \frac{N_f D(K)}{2\bar{K}K}(\bar{K}-K)-\frac{D(
K) }{K}+\frac{\partial D(K) }{\partial K}\right)~dt\\+ \sqrt{%
2D(K) }~dW,
\end{multline}
which can be used to generate the correct canonical distribution. This
result is independent on the choice of the function $D(K)$, but different
choices can lead to different speeds of equilibration. Here we choose
\begin{equation}
D(K)=\frac{2K\bar{K}}{N_{f}\tau },
\end{equation}%
where the arbitrary parameter $\tau $ has the dimension of a time and
determines the time-scale of the thermostat such as in Berendsen's
formulation. This leads to a very
transparent expression for the auxiliary dynamics
\begin{equation}
dK=(\bar{K}-K)\frac{dt}{\tau }+2\sqrt{\frac{K\bar{K}}{N_{f}}}\frac{dW}{\sqrt{%
\tau }}.  \label{stochastic-evolution}
\end{equation}%
Without the stochastic term this equation reduces to that of the standard
thermostat of Berendsen. In the limit $\tau=0$, the stochastic evolution is
instantly thermalized and this algorithm reduces exactly to the stochastic
velocity-rescaling approach described in Section~\ref{sec-rescale}.
On the other hand, for $\tau \rightarrow \infty $,
the Hamiltonian dynamics is recovered. When a system
is far from equilibrium, the deterministic part in Eq.~(\ref{stochastic-evolution})
dominates and our algorithm leads to fast equilibration like the Berendsen's
thermostat. Once the equilibrium is reached the proper canonical ensemble is sampled,
at variance with the Berendsen's thermostat.

There is no need to apply additional self-consistency procedures to enforce
rigid bond constraints, as in the case of Andersen's thermostat, 
since the choice of a single rescaling factor for
all the atoms automatically preserves bond lengths. Furthermore, the total linear
momentum and, for non-periodic systems, the angular momentum are conserved. The
formalism can also be trivially extended to thermalize independently different
parts of the system, e.g., solute and solvent, even
using different parameters $\tau$ for the different subsystems.
Interestingly, dissipative particle dynamics\cite{sodd+03pre}
can be included in our scheme if different
thermostats are applied to all the particles pairs that are within a given
distance.

We have already noted that Berendsen's thermostat can be recovered from ours by
switching off the noise. Also Nos{\'e}'s thermostat can be recast in a form
that parallels our formulation. To this effect, it is convenient to rewrite
the auxiliary variable $\xi$ of the Nos{\'e}-Hoover thermostat in adimensional
form  $\xi =\frac{\zeta }{\tau }$ and the mass of the thermostat as $\frac{%
N_{f}}{\beta }\tau ^{2}$.
In our scheme, Nos{\'e}-Hoover dynamics is obtained through these
auxiliary
equations for $\zeta $ and $K$:
\begin{subequations}
\begin{equation}
dK=-2\zeta K\frac{dt}{\tau },
\end{equation}%
\begin{equation}
d\zeta =\left( \frac{K}{\bar{K}}-1\right) \frac{dt}{\tau }.
\end{equation}
\end{subequations}
The corresponding Liouville equation for the probability distribution
$P(K,\zeta)$ is 
\begin{multline}
\tau \frac{\partial P(K,\zeta ;t)}{\partial t}=2\zeta P+2\zeta K\frac{%
\partial P(K,\zeta ;t)}{\partial K}\\ -
\left( \frac{K}{\bar{K}}-1\right) \frac{%
\partial P(K,\zeta ;t)}{\partial \zeta }
\end{multline}
which is stationary for
\begin{equation}
\bar{P}(K,\zeta )~dKd\zeta\propto K^{\left(\frac{N_{f}}{2}-1\right)}e^{-\beta K}e^{-\frac{N_{f}\zeta
^{2}}{2}}
~dKd\zeta
\end{equation}%
which is the desired distribution. By comparing this formulation of the
Nos{\'e}-Hoover thermostat with our scheme we see that the variable $\zeta$
plays the same role as the noise. In the Nos{\'e}-Hoover scheme
the chaotic nature of the coupled equations of motion leads to a
stochastic $\zeta$.
When the system to be thermostated is poorly ergodic,
$\zeta$ is no longer stochastic and a chain of thermostats
is needed.\cite{mart92jcp}

\subsection{Controlling the integration time-step}
\label{sec-checking-time-step}

When integrating the equations of motion using a finite time-step, a
technical but important issue is the choice of an optimal value for the
time-step. The usual paradigm is to check if the constants of motion are
properly conserved. For example, in the microcanonical ensemble, sampled
using the Hamilton's equations, the check is done on the total energy of the
system which is given by the Hamiltonian $H(x)$, where $x=\left( p,q\right) $
is a point in phase space. When the Nos\'{e}-Hoover thermostat is used, the
expression for the conserved quantity $H_{\text{Nos{\'e}}}$ is more complex and can be
recast in the form: 
\begin{equation}
H_{\text{Nos{\'e}}}=H(x)+\frac{N_{f}}{\beta }\left( \frac{\zeta ^{2}}{2}+\int_{0}^{t}%
\frac{dt^{\prime }}{\tau }\zeta \left( t^{\prime }\right) \right) .
\end{equation}
In this section we propose a quantity that can play the same role for our
thermostat, even though we are dealing with a stochastic
process.

Let us consider a deterministic or stochastic dynamics aimed at sampling a
given probability distribution $\bar{P}(x)$.
It is convenient to consider a discrete form of dynamics. This is not a
major restriction and, after all, on the computer any dynamics is implemented
as a discrete process. Starting from a point $x_{0}$ in the phase space we
want to generate a sequence of points $x_{1},x_{2},\dots$ distributed
according to a probability as close as possible to $\bar{P}(x)$.
Let $M(x_{i+1}\leftarrow x_{i})~dx_{i+1}$ be the conditional probability of reaching the point
$x_{i+1}$ given that the system is at $x_{i}$. In order to
calculate statistical averages which are correct independently
of $M$, each point visited has to be weighted by
$w_{i}$ which measures the probability that $x_{i}$ is in the target
ensemble. The ratio between the weights of successive points is
\begin{equation}
\frac{w_{i+1}}{w_{i}}=\frac{M(x_{i}^*\leftarrow x^*_{i+1})\bar{P}(x^*_{i+1})}{%
M(x_{i+1}\leftarrow x_{i})\bar{P}(x_{i})},
\label{eq-weights}
\end{equation}
where the conjugated point $x_{i}^{\ast }$ is obtained from $x_{i}$ inverting
the momenta, i.e., if $x=(p,q)$, $x^{\ast }=(-p,q)$. If the
dynamics exactly satisfies the detailed balance one must have
$\frac{w_{i+1}}{w_{i}}=1$, which implies that~$w$ must be
constant. However, if $\bar{P}(x)$ is sampled in an approximated way, the
degree to which $w$ is constant can be used to assess the accuracy of the
sampling.

Rather than in terms of weights, it is convenient to express this principle in terms
of an effective energy
\begin{equation}
\tilde{H}_{i}=-\frac{1}{\beta }\ln w_{i}.  \label{eq-def-htilde}
\end{equation}
The evolution of $\tilde{H}$ is given by
\begin{multline}
\tilde{H}_{i+1}-\tilde{H}_{i}=-\frac{1}{\beta }\ln
\left(\frac%
{M(x_{i}^*\leftarrow x_{i+1}^*)\bar{P}(x_{i+1}^*)}%
{M(x_{i+1}\leftarrow x_{i})\bar{P}(x_i)}
\right)
\label{eq-htilde} \\
=-\frac{1}{\beta }\ln
\left(\frac%
{M(x_{i}^*\leftarrow x_{i+1}^*)}%
{M(x_{i+1}\leftarrow x_{i})}
\right) +H( x_{i+1})
-H( x_{i}),
\end{multline}%
where the last line follows in the case of a canonical distribution
$\bar{P}(x)\varpropto e^{-\beta H\left( x\right) }$.

\begin{figure}
\includegraphics[clip,width=0.45\textwidth]{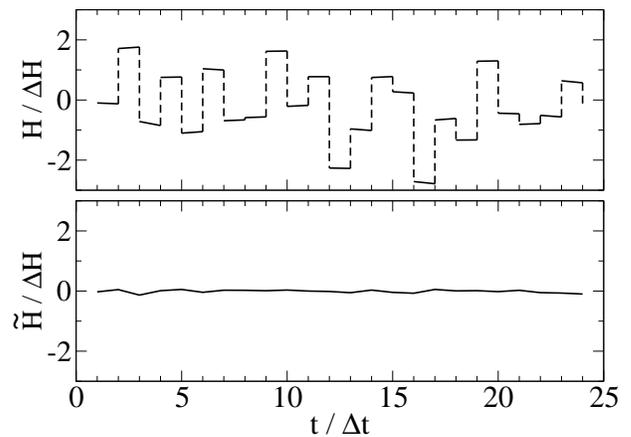}
\caption{
Schematic time series for $H$ (upper) and $\tilde{H}$ (lower),
in units of the root-mean-square fluctuation
of $H$. Time is in unit of the integration time-step.
The solid lines represent the increments due to the velocity Verlet
step, which is almost energy preserving.
The dashed lines represent the increments due to the velocity-rescaling.
Since only the changes due to the velocity Verlet steps are accumulated into
$\tilde{H}$, this quantity is almost constant. On the other hand,
$H$ has the proper distribution.
}
\label{fig-scheme}
\end{figure}

Let us now make use of this result in the context of our dynamics. This is
most conveniently achieved if we solve the equations of motion alternating
two steps. One is a velocity Verlet step, or any other area-preserving and
time-reversible integration algorithm. In a step with such property,
$M(x_{i}^*\leftarrow x_{i+1}^*)=M(x_{i+1}\leftarrow x_{i})$
and the change in $\tilde{H}$ is equal to the change in $H$. The other is a
velocity-rescaling step in which the scaling factor is determined via Eq.~(\ref{stochastic-evolution}).
If we use the exact solution of Eq.~(\ref{stochastic-evolution})
derived in the Appendix, this step satisfies the detailed balance 
and therefore does not change $\tilde{H}$.
An idealized but realistic example of time evolution of $H$ and $\tilde{H}$
is shown in Fig.~\ref{fig-scheme}.

If we use this analysis and go to the limit of an infinitesimal time-step we find
\begin{equation}\label{eq-conserved-quantity}
\tilde{H}(t)=H(t)-\int_{0}^{t}(\bar{K}-K(t'))\frac{dt'}{\tau }-2\int_{0}^{t}\sqrt{\frac{%
K(t')\bar{K}}{N_{f}}}\frac{dW(t')}{\sqrt{\tau }},
\end{equation}%
where the last two terms come from the integration of Eq.~(\ref{stochastic-evolution}) along the trajectory. Note
that a similar integration along the path is present in $H_{\text{Nos{\'e}}}$. However in our scheme
a stochastic integration is also necessary. In the continuum limit the
changes in energy induced by the rescaling compensate exactly the fluctuations
in $H$. For a finite time-step this compensation is only
approximate and the conservation of $\tilde{H}$ provides a measure on the
accuracy of the integration. This accuracy has to be interpreted in the sense of the ability
of generating configurations representative of the ensemble.
The physical meaning of Eq.~(\ref{eq-conserved-quantity}) is that the fluxes of energy between
the system and the thermostat are exactly balanced.

A further use of $\tilde{H}$ is possible whenever high accuracy results are
needed and even the small error deriving from the use of a finite time-step
integration needs to be eliminated.
In practice, one can correct this error reweighting\cite{wong-lian97pnas} the points
with $w_{i}\varpropto e^{-\beta \tilde{H}_i}$.
Alternatively, segments of
trajectories can be used in a hybrid Monte Carlo scheme\cite{duan+97plb} to generate new
configurations which are accepted or rejected with probability $\min
(1,e^{-\beta (\Delta \tilde{H})})$.

From the discussion above one understands that in many ways $\tilde{H}$ has a
role similar to $E$ in the microcanonical ensemble.
It is however deeply different: while in the microcanonical ensemble
$E$ defines the ensemble and has a physical meaning,
in the canonical ensemble the value of $\tilde{H}$ simply depends on the
chosen initial condition. Thus, the value of $\tilde{H}$ can be only compared
for points belonging to the same trajectory.

\section{Applications}
\label{sec-applications}

In this section we present a number of test applications of our
thermostating procedure. Moreover, we compare its properties with those of
commonly adopted thermostats, such as the Nos\'e-Hoover and the Berendsen
thermostat. To test the efficiency of our
thermostat we compute the energy fluctuations and the dynamic properties
of two model systems, namely a Lennard-Jones system and water, in their crystalline
and liquid phases.
All the simulations have been performed using a modified version of the DL POLY code\cite{dlpoly,dlpoly2}.

We adopt the parameterization of the Lennard-Jones potential for argon,
and we simulate a cubic box containing 256 atoms.
Calculations have been performed on the 
crystalline solid fcc phase at a temperature of 20 K, and on the 
liquid phase at 120 K. The cell side is 21.6 \AA{} for the solid
and 22.5 \AA{} for the liquid.
Water is modeled through the commonly used TIP4P potential:\cite{tip4p}
water molecules are treated as rigid bodies and interact {\sl via} the dispersion forces
and the electrostatic potential generated by point charges. The long-range electrostatic
interactions are treated by the particle mesh Ewald method.\cite{dard+93jcp}
The energy fluctuations and the dynamic properties, such as the frequency spectrum
and the diffusion coefficient, have been computed on models of liquid water and
hexagonal ice I$_h$, in cells containing 360 water molecules with periodic
boundary conditions. The model of ice I$_h$, with a fixed density of 0.96 g/cm$^3$,
has been equilibrated at 120 K, while the liquid has a density of 0.99 g/cm$^3$ and
is kept at 300 K.

\subsection{Controlling the integration time-step}

\begin{figure}
\includegraphics[clip,width=0.50\textwidth]{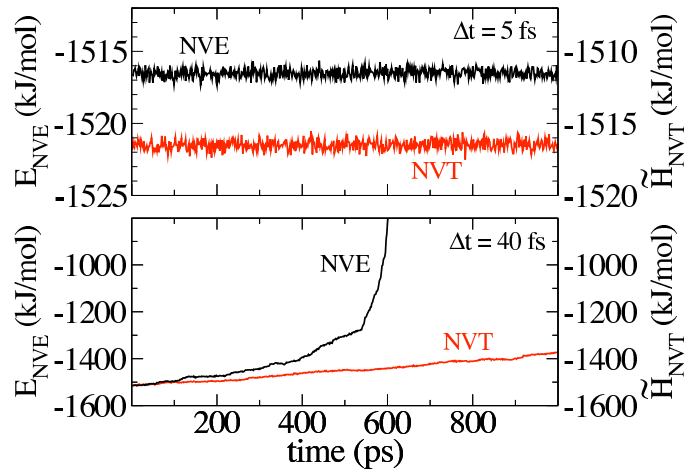}
\caption{(color online) Total energy $E$ (left axis) and effective energy $\tilde{H}$ (right axis),
respectively for a NVE simulation and for a NVT simulation using
our thermostat, with $\tau=0.1$ ps, for Lennard-Jones at 120 K.
In the upper panel, the calculation is performed with a time-step $\Delta t=5$ fs,
and $E$ (or $\tilde{H}$) does not drift. In the lower panel, the
calculation is performed with a time-step $\Delta t=40$ fs,
and $E$ (or $\tilde{H}$) drifts.
}
\label{fig-htilde-lj}
\end{figure}

As discussed in Section~\ref{sec-checking-time-step}, the effective energy
$\tilde{H}$ can be used to verify the sampling accuracy and plays a role
similar to the total energy in the microcanonical ensemble.
In Fig.~\ref{fig-htilde-lj} we show the time-evolution of $\tilde{H}$
for the Lennard-Jones system at 120 K with two different integration time-steps,
namely $\Delta t=5$ fs and $\Delta t=40$ fs. In both cases,
we use $\tau=0.1$ ps for the thermostat time-scale.
With $\Delta t=5$ fs the integration is accurate and the
effective energy $\tilde{H}$ is properly conserved, in the sense that it
does not exhibit a drift. Moreover, its fluctuations are rather small, approximately
$0.3$ kJ/mol:
for a comparison, the root-mean-square fluctuations in $H$ are on the order of 
$16$ kJ/mol.
When the time-step is increased to $\Delta t=40$ fs, the integration
is not accurate and there is a systematic drift in the effective energy.
For a comparison we show also the time-evolution of $E$ in a
conventional NVE calculation. The fluctuations and drifts
for $E$ in the NVE calculation are similar to the fluctuations and drifts
for $\tilde{H}$ in the NVT calculation.
We notice that, while in the NVE calculation the 
system explodes at some point, the NVT calculation is always
stable thanks to the thermostat. However, in spite of the stability,
the drift in $\tilde{H}$ indicates that
the sampling is inaccurate under these conditions.

\subsection{Energy fluctuations}

\begin{figure}
\includegraphics[clip,width=0.45\textwidth]{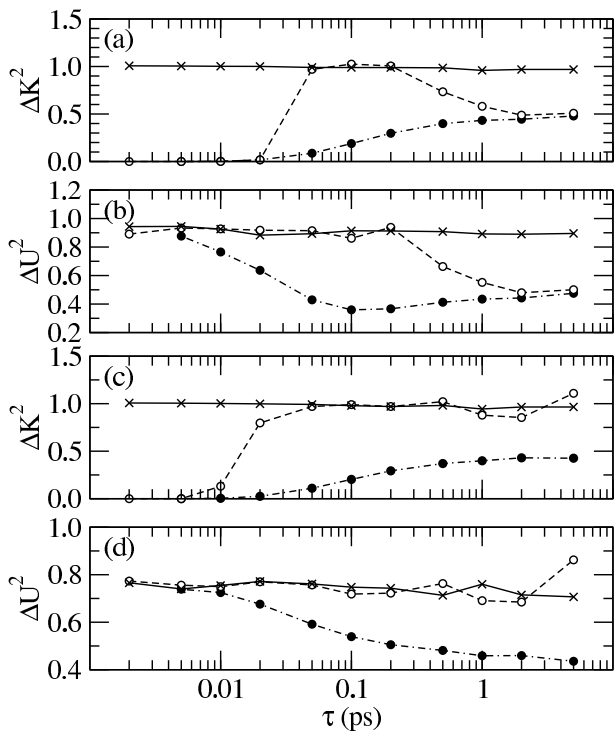}
\caption{Square fluctuations of the kinetic energy $\Delta{K}^2$ and of the
potential energy $\Delta U^2$, in units of $N_fk_B^2T^2/2$,
for a Lennard-Jones solid at 20 K [panels (a) and (b)]
and liquid at 120 K [panels (c) and (d)], using
the Berendsen ($\bullet$, dashed-dotted), the Nos\'e-Hoover ($\circ$, dashed) and our ($\times$, solid) thermostat,
plotted as function of the characteristic time of the thermostat $\tau$.
In these units the analytical value for the fluctuations of the kinetic energy is 1.}
\label{fig-lj-fluct}
\end{figure}

\begin{figure}
\includegraphics[clip,width=0.45\textwidth]{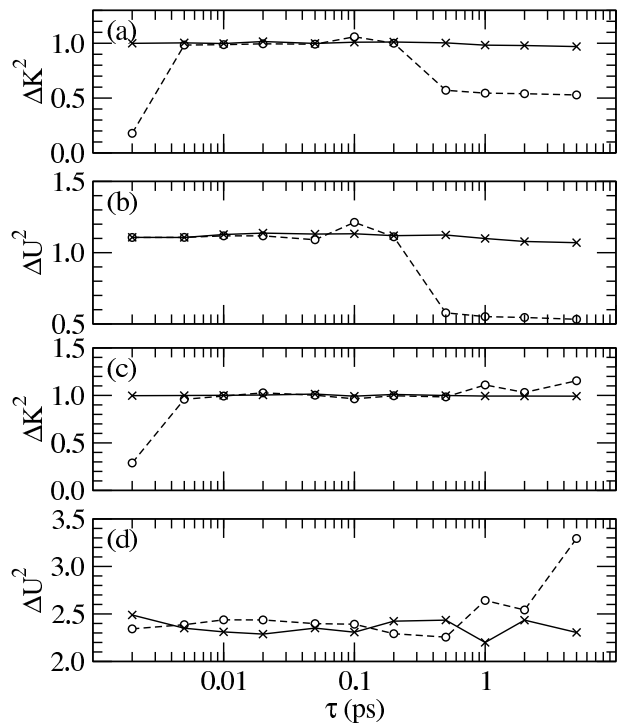}
\caption{Square fluctuations of the kinetic energy $\Delta{K}^2$ and of the
potential energy $\Delta U^2$, in units of $N_fk_B^2T^2/2$,
for ice at 120 K [panels (a) and (b)]
and water at 300 K [panels (c) and (d)], using
the Nos\'e-Hoover ($\circ$, dashed) and our ($\times$, solid) thermostat,
plotted as function of the characteristic time of the thermostat $\tau$.
In these units the analytical value for the fluctuations of the kinetic energy is 1.}
\label{fig-h2o-fluct}
\end{figure}

While the average properties are equivalent in all the ensembles, the
fluctuations are different. Thus we use the square fluctuations
of the configurational and kinetic energies, which are related to the
specific heat of the system,\cite{alle-tild87book} to check
whether our algorithm samples the canonical ensemble.
Therefore we perform 1 ns long molecular dynamics runs using our thermostat with
different choices of the parameter $\tau$, spanning three
orders of magnitude. 
An integration time-step $\Delta t=5$ fs is adopted, which yields a satisfactory 
conservation of the effective energy, as verified in the previous section.
For comparison,
we also calculate the fluctuations using the Nos\'e-Hoover thermostat,
which is supposed to sample the proper ensemble.
The results for the Lennard-Jones system, both solid and liquid, are presented in
Fig.~\ref{fig-lj-fluct}.
The fluctuations are plotted in units of the ideal-gas kinetic-energy fluctuation.
For the liquid [panels (c) and (d)] the Nos\'e-Hoover
and our thermostat give consistent results for a wide range of values of
$\tau$ for both the kinetic and configurational energy fluctuations.
The Nos\'e-Hoover begins to fail only in the regime of small $\tau$,
due to the way the extra variable of the thermostat is integrated.
For the solid [panels (a) and (b)] the ergodicity problems
of Nos\'e-Hoover's thermostat appear for $\tau>0.2$ ps, in terms of poor sampling.
We notice that increasing $\tau$ the fluctuations
tend to their value in the microcanonical ensemble.
On the other hand, our procedure is correct over the whole $\tau$ range,
both for the solid and for the liquid.
In Fig.~\ref{fig-lj-fluct} we also plot the fluctuations calculated
using the Berendsen thermostat. It is clear that both for the liquid
and for the solid the results are strongly dependent on the choice of
the $\tau$ parameter. In the 
limit $\tau\rightarrow 0$ the Berendsen thermostat tends to the isokinetic
ensemble, which is consistent with the canonical one for
properties depending only on configurations:\cite{evan-morr84cpc}
thus, the fluctuations in the configurational energy tend to the
canonical limit, while the fluctuations in the kinetic energy
tend to zero.
In Fig.~\ref{fig-h2o-fluct} we present a similar analysis done on
water [panels (c) and (d)] and ice [panels (a) and (b)].
For this system the equations of motion have been integrated with
a time-step $\Delta t=1$ fs.
In this case we only performed the calculation with Nos\'e-Hoover's thermostat
and with ours.
Nos\'e-Hoover's thermostat is not efficient for ice in the case of $\tau$ larger than 0.2 ps,
and also for water in the case of $\tau$ larger than 2 ps.
On the other hand, the performance of our thermostat
is again fairly independent from the choice of $\tau$.

\subsection{Dynamic properties}

\begin{figure}
\includegraphics[clip,width=0.45\textwidth]{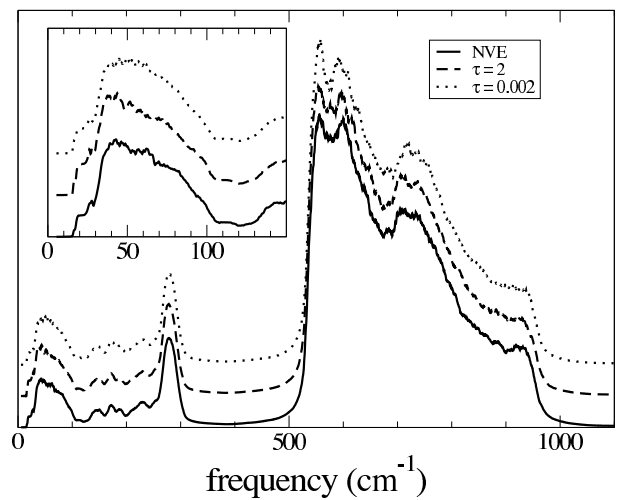}
\caption{
Vibrational density of states for the hydrogen atom in ice I$_h$
at 120 K. The spectra obtained with different values of the relaxation
time $\tau$ of our thermostat (dashed and dotted lines) are compared to a simulation
in the NVE ensemble (solid line).
}
\label{vibrational}
              
\end{figure}

In order to check to what extent our thermostat affects
the dynamic properties of the systems, we have computed the vibrational
spectrum of the hydrogen atoms in ice I$_h$ from the Fourier transform of 
the velocity-velocity autocorrelation function. 
The spectra have been computed, sampling
100 ps long trajectories in the NVT ensemble every 2 fs,
at a temperature of 120 K, with two different values of the relaxation time $\tau$
of the thermostat (see Fig. \ref{vibrational}). 
They are compared to the spectrum of frequencies obtained in a run in the
microcanonical ensemble. In all these runs the integration time-step $\Delta t$ has been reduced to 0.5 fs.
Two main regions can be distinguished in the vibrational spectrum of ice: 
a low frequency band corresponding to the translational modes (on the left in Fig. 
\ref{vibrational}) and a band at higher frequency related to the librational modes. 
Due to the use of a rigid model the high frequency intramolecular modes
are irrelevant.
All the features of the vibrational 
spectrum of ice I$_h$ are preserved when our thermostat is used.
When compared with the spectrum obtained from the NVE simulation,
no shift of the frequency of the main peaks 
is observed for $\tau=2$ ps and the changes in their intensities are within numerical errors. 
It is worth noting that, although the thermostat acts directly on the particle 
velocities, it does not induce the appearance of fictitious peaks in the spectrum.
The simulation done with $\tau=0.002$ ps shows that with
a very short $\tau$ the shape of the first translational broad peak is
slighly affected (see the inset in Fig.~\ref{vibrational}).

\begin{table}
\begin{center}
\begin{tabular}{cc}
$\tau$ (ps)  &  D ($10^{-5}$cm$^2$/s) \\
\hline
0.002  &  3.63 $\pm$ 0.01  \\
0.02   &  3.44 $\pm$ 0.06  \\
0.2    &  3.51 $\pm$ 0.05  \\
2.     &  3.53 $\pm$ 0.01  \\
\hline\hline
NVE    &  3.47 $\pm$ 0.03  \\
\end{tabular}
\caption{The diffusion coefficient $D$ of water at 300 K, as a function of the
relaxation time $\tau$ of the thermostat. For a comparison, also
the value obtained from a NVE trajectory is shown.}
\label{diffusion}
\end{center}                                           
\end{table}

The performances of our thermostat have been tested also with respect to the
dynamic properties of liquids, computing the self-diffusion coefficient $D$ of TIP4P 
water. $D$ is computed from the mean square displacement, through Einstein's relation,
on 100 ps long simulations equilibrated at a temperature of 300 K. 
The results for different values of $\tau$ are reported in
Tab.~\ref{diffusion} and compared to the value of $D$ extracted from an NVE simulation.
The results obtained applying our thermostat are compatible with the values 
of $D$ reported in literature for the same model of water,\cite{vdspoel98} and are 
consistent with the one extracted from the microcanonical run. A marginal
variation with respect to the reference value occurs for very small $\tau=0.002$ ps.

\section{Conclusion}
\label{sect-conclusion}

We devised a new thermostat aimed at performing molecular dynamics
simulations in the canonical ensemble. This scheme is derived from a
modification of the standard velocity-rescaling with a properly
chosen random factor, and generalized to a smoother formulation
which resembles the Berendsen thermostat.
Under the assumption of ergodicity,
we proved analytically that our thermostat 
samples the canonical ensemble.
Through a proper combination with a barostat, it can be used to sample
the constant pressure--constant temperature ensemble.
We check the ergodicity assumption on realistic systems and we compare the ergodicity
of our procedure with that of the Nos\'e-Hoover thermostat, finding
our method to be more ergodic. 
We also use the concept of sampling accuracy rather than trajectory accuracy
to assess the quality of the numerical integration of our scheme.
To this aim, we introduce a new quantity, which we dub effective
energy, which measures the ensemble violation.
This formalism allows a robust check on the finite time-step
errors and can be easily extended to other kinds of
stochastic molecular dynamics, such as the Langevin dynamics.

\section{Acknowledgements}

The authors would like to thank Davide Branduardi for
useful discussions. Also Gabriele Petraglio is acknowledged
for carefully reading the manuscript.

\appendix

\section{Exact propagator for the kinetic energy}

For the thermostat designed here it is essential that the exact solution of
Eq.~(\ref{stochastic-evolution}) is used.
In the search for an analytical solution we are inspired by the fact that, if each individual momentum
is evolved using a Langevin equation, then the evolution of $K$ is described by the same
Eq.~(\ref{stochastic-evolution}). Thus we first  define an auxiliary set of $N_f$ stochastic processes $%
x_{i}(t)$ with the following equation of motion 
\begin{equation}
dx_{i}(t)=-\frac{x_{i}(t)}{2}~\frac{dt}{\tau}+\frac{dW_{i}(t)}{\sqrt{\tau }}.
\label{eq-motion-x}
\end{equation}%
This is the equation for an overdamped harmonic oscillator which is known
also as the Ornstein-Uhlenbeck processes for which  an analytical solution
exists:\cite{risk89book}
\begin{equation}
x_{i}(t)=x_{i}(0)e^{-\frac{t}{2\tau }}+\sqrt{1-e^{-\frac{t}{\tau }}}R_{i},
\label{eq-propagation-x}
\end{equation}%
where the $R_{i}$'s are independent random numbers from a Gaussian distribution
with unitary variance. Then we define the variable  $y$
\begin{equation}
y(t)=\frac{1}{N_f}\sum_{i=1}^{N_f}x_{i}^{2}(t).\label{eq-def-y}
\end{equation}%
The equation of motion for $y$ is obtained applying the Itoh rules\cite{gard03book} to Eq.~%
(\ref{eq-motion-x}) and recalling that the increments $dW_{i}$ are
independent
\begin{equation}
dy(t)=(1-y(t))\frac{dt}{\tau }+2\sqrt{\frac{y(t)}{N_f}}\frac{dW(t)}{\sqrt{\tau 
}}.  \label{eq-motion-y}
\end{equation}%
Since the equation of motion for $x$ is invariant under
rotation, we can assume without loss of generality that
at $t=0$ the
multidimensional vector $\left\{ x_{i}\right\} $ is oriented along its
first component
\begin{equation}
\left\{ x_{i}(0)\right\} =\left\{ \sqrt{N_fy(0)},0,...\right\}.
\end{equation}
Combining Eqs.~(\ref{eq-def-y}) and (\ref{eq-propagation-x}) we obtain for $y(t)$  at
finite time
\begin{multline}
y(t)=e^{-t/\tau}y(0) + (1-e^{-t/\tau})\sum_{i=1}^{N_f}\frac{R_i^2}{N_f} \\
+2e^{-t/2\tau}\sqrt{1-e^{-t/\tau}}\sqrt{\frac{y(0)}{N_f}}R_1.
\end{multline}%
We now observe that with the substitution $y(t)=\frac{K_{t}}{\bar{K}}$
Eq.~(\ref{eq-motion-y}) is equivalent to Eq.~(\ref{stochastic-evolution}).
Thus, with simple algebra, we find the desired expression for the rescaling
factor,
\begin{multline}
\label{eq-propagation-y2}
\alpha^2 = e^{-\Delta t/\tau} +
\frac{\bar{K}}{N_fK} (1-e^{-\Delta t/\tau})
(
R_1^2+
\sum_{i=2}^{N_f} R_i^2
) \\
+2e^{-\Delta t/2\tau}\sqrt{\frac{\bar{K}}{N_fK} (1-e^{-\Delta t/\tau})} R_1
\end{multline}%
We observe here that there is no need to draw all the $R_i$'s Gaussian numbers, because $%
\sum_{i=2}^{N_f}R_{i}^{2}$ can be drawn directly from the Gamma distribution $p_{\frac{%
N_{f}-1}{2}}(x) =\frac{x^{\left(\frac{N_{f}-1}{2}-1\right)}e^{-x}}{\Gamma \left( 
\frac{N_{f}-1}{2}\right) }$ if $N_{f}-1$ is even or by adding a squared random Gaussian
number to that extracted from $p_{\frac{N_{f}-2}{2}}(x) =%
\frac{x^{\left(\frac{N_{f}-2}{2}-1\right)}e^{-x}}{\Gamma \left( \frac{N_{f}-2}{2}\right) }
$ if $N_{f}-1$ is odd.\cite{nr}
A routine that evaluates Eq.~(\ref{eq-propagation-y2}) is available upon request.

\end{document}